# Immortality as a Physical Problem

Mark Ya. Azbel'

School of Physics and Astronomy, Tel-Aviv University,

Ramat Aviv, 69978 Tel Aviv, Israel;

Max-Planck-Institute für Festkorperforschung – CNRS,

F38042 Grenoble Cedex 9, France

## Abstract

Well protected human and laboratory animal populations with abundant resources are evolutionary unprecedented. Physical approach, which takes advantage of their extensively quantified mortality, establishes that its dominant fraction yields the exact law, whose universality from yeast to humans is unprecedented, and suggests its unusual mechanism. Singularities of the law demonstrate new kind of stepwise adaptation. The law proves that universal mortality is an evolutionary byproduct, which at any age is reversible, independent of previous life history, and may be disposable. Recent experiments verify these predictions. Life expectancy may be extended, arguably to immortality, by relatively small and universal biological amendments in the animals. Indeed, it doubled with improving conditions in humans; increased 2.4-fold with genotype change in Drosophila, and 6-fold (to 430 years in human terms), with no apparent loss in health and vitality, in nematodes with a small number of perturbed genes and tissues. The law suggests a physical mechanism of the universal mortality and its regulation.

**1. Immortality and physics.** Every new field in physics introduced unanticipated concepts and laws. Even thermodynamics of classical particles with reversible mechanics yielded irreversibility. However, biophysicists reduce complex live systems to conventional models. Consider an alternative approach.

Biological diversity evolved in evolutionary selection of the fittest via death of the frail. In the wild competition for sparse resources is fierce, and only relatively few genetically fittest animals survive to their evolutionary "goal"- reproduction. There are no evolutionary benefits from genetically programmed death of very few survivors significantly beyond reproductive age. Well protected human and laboratory animal populations with abundant resources are evolutionary unprecedented and "unanticipated". Their accidental extrinsic mortality is low. Living conditions, phenotypes, genotypes, tissues of laboratory animals may be manipulated [1-5] and become evolutionary unprecedented. Thus, "protected" mortality may be biologically unusual and calls for an alternative study. Human mortality is well quantified. Its knowledge is crucial for economics, taxation, insurance, etc. Its study was started in 1693 by Halley (of the Halley comet) and followed in 1760 by Euler [6]. In developed countries human birth and death are accurately registered for well over a century. By now demographers accumulated millions of highly reliable mortality data [7]. Biodemographers collected them for animals from yeast to mammals (see, e. g., [8]). Extensive and readily available data allow one to establish, with approach characteristic of physics rather than of biology or demography, the exact universal law for a dominant fraction of mortality in protected populations. The law reduces a multitude of hardly quantifiable environmental and population factors (in humans they are uncontrollably non-stationary and heterogeneous) to few specified parameters. It is the same for all animals, i.e. invariant to any transformations from single cell yeast to humans. Such "general invariance" is inconsistent with common biological wisdom, yet agrees with experiments (see later). The law, which was preserved in an entire animal evolution, is its conservation law. Contrary to all existing theories [9] and models [10], it predicts rapid (compared to the life span) and stepwise adaptation to environment, a possibility of mortality reversal to younger age, of drastic increase in life expectancy, and even immortality. Recent experiments are consistent with predictions (which were first made empirically [11])-see section 3. Most remarkably, changes in small number of genes and tissues in nematodes [5] increase their mean life span six-fold to 124 days (430 years in human terms) with no apparent loss in health and vitality. Universality of the law allows one to study its physical mechanism and the ways to regulate it in the simplest case of, e.g., yeast. Physical experiments may relate this mechanism to processes in and genetics of a cell and yield a microscopic model of unprecedented "general invariance".

**2. Universal law.** Demographic life tables present millions of accumulated mortality data in different countries over their history (see, e.g., [7]). The data list, in particular, "period" probabilities $q(x)$ (for survivors to x years) and $d(x)$ (for live newborns) to die between the ages x and (x+1) [note that $d(0)=q(0)$]; the probability $l(x)$ to survive to x for live newborns; the life expectancy $e(x)$ at the age x for males and females who died in a given country and calendar year. The tables also present [7] the data and procedures which allow one to calculate "cohort" probabilities for those who were born in a given calendar year. To estimate and forecast period mortality, demographers developed over 15 approximations [12]. Biodemographers similarly approximated animal mortality [8]. Populations, their conditions and heterogeneity are different, yet each approximation reduces period mortality of any given population to few parameters [13]. So, I conjecture that under certain conditions a dominant fraction of heterogeneous mortality in all populations is accurately related to a specific number of parameters. Such conjecture is sufficient to derive this universal relation. Chose "additive" mortality variables, which are the averages of their values in different population groups of the same age. If the population consists of the groups with the number $N^G(x)$ of survivors to age $x$ (in years for humans, days for flies, etc) in a group $G$, then the total number of survivors $N(x)$ is the sum of $N^G(x)$ over all $G$. If $c_G$ is the ratio of the population and $\ell^G(x)$ is the survivability to $x$ in the group $G$, then the probability $l(x)$ to survive to $x$ for live newborns is $l(x) = N(x)/N(0) = \Sigma\, c_G\, \ell^G(x) = <\Sigma \ell^G(x)>$, i.e. the average of $\ell^G(x)$ over all groups. The most age specific additive variable is $d(x) = \ell(x) - \ell(x+1)$. The most time specific additive variable is $d(0)$ which may depend on the time from conception to x=1 only. Since the probability to die at the age x is $q(x) = \ell(x)d(x)$ and $l(0) = 1$, so "infant mortality" $q(0) = d(0)$. In the simplest case (which may easily be generalized) of one variable, universality implies that the relation between d($x$) and $q(0)$ (here and on d and q denote the fractions which yield the universal law) is the same as the relation between their values in any of the groups in the interval. So, if $d(x) = f_x[q(0)]$; then $d^G(x) = f_x[q^G(0)]$. Since additive $d(x) = <d^G(x)>$, $q(0) = <q^G(0)>$, so $<f_x[q^G(0)]> = f_x(<q^G(0)>)$. According to a simple property of stochastic variables, if the average of an analytical function is equal to the function of the average, then the function is linear. In a general case the (singular) function is piecewise linear: in the j-th interval (denote its population as an "echelon"), $j = 1, 2, …, J$,

$$d(x) = d_j(x) = a_j(x)q(0) + b_j(x) \text{ when } q_j < q^G(0) < q_{j+1} \quad (1)$$

When infant mortality $q(0)$ of an echelon reaches its boundary, it homogenizes. Since, by Eq. (1), d($x$) at all ages reduce to infant mortality, they simultaneously reach the interval boundary and, together with q(0), homogenize there. (Two such "ultimate" boundaries are well known: $q(x) = 0$ implies that nobody dies at, and $\ell(x) = 0$ implies that nobody survives beyond, the age $x$). At any age $d_j(x) = d_{j+1}(x)$ when $q(0) = q_{j+1}$. This reduces all $d_j(x)$ to $(J+1)$ universal functions of $x$ and $(J-1)$ universal constants. In a special case, when every universal population is homogeneous (i.e. when J is infinite, and $q_j \rightarrow q_{j+1}$), $d(x)$ vs $q(0)$ may be arbitrary. Experiments are consistent with finite $J$ only (see the next section).

Consider an arbitrarily heterogeneous population. Suppose its fractions and the fractions of its infant mortality $q(0)$ in the j-th echelon are correspondingly $c_j$ and $f_j = c_j q_j(0)/q(0)$. Then $d(x) = \Sigma c_j d_j(x)$ reduces to the universal dependence on these parameters and $q(0)$:

$$d(x) = aq(0) + b; \quad a = \Sigma f_j a_j; \quad b = \Sigma c_j b_j ; \tag{2}$$

where $0 < c_j, f_j < 1$; $\Sigma c_j = \Sigma f_j = 1$. (Here and on the argument x is skipped in a and b).

By Eq. (2), in a general case the universal law reduces mortality to population specific parameters $c_j$, $f_j$ and $q(0)$; to species specific constants $q_j$ and to universal (when properly scaled with $q_j$) functions $a_j$, $b_j$ of age x.

Equations (1, 2) map on phase equilibrium. Present Eq. (1) as

$$d(x) = Cd^{(j)}(x) + (1-C)d^{(j+1)}(x), \tag{3}$$

where

$$d^{(j)}(x) = a_j(x)q^{(j)} + b_j(x) \tag{4}$$

Then $d^{(j)}(x)$ may be interpreted as the universal "equation of state" of the $j$-th "phase" and C as its "concentration" determined by $q(0) = d(0)$. An echelon reduces to two, an arbitrary population to $(J+1)$ phases [13].

The universal law was first established empirically [9], and its extrapolation to $q(0)=0$ yielded human $l(80)=0$ (Figs 1 and 4 in the successive refs. 11; see also later), thus $q(x)=0$ and, by Eqs. (1,2), $b_1(x)=0$ till x=80. In virtue of universality, this implies, also in contrast to all theories [9] and models [10], zero universal mortality till certain old age in all animals. Higher accuracy may change these "empirical" zeroes into small universal values. Alternatively, if $b_1(x)=0$ till certain age $x=x^*$, then, according to a well known mathematical theorem, either $b_1(x)$ has a singularity at $x=x^*$, or $b_1(x)=0$ at any age. The latter case implies a possibility of ultimate immortality.

**3. Verification.** Remarkably, at certain ages the most unusual prediction of zero mortality may have already been observed, but not appreciated. In 2001 Switzerland [7] in each age group (with~60,000 girls) only 1 girl died at 5, 9, and 10 years; 5 from 4 to 7 and from 9 to 13 years; 10 or less from 2 till 17 years; no more than 16 from 2 till 26. Statistics is similar in all 1999-2002 Western developed countries [7]. Such low values of a stochastic quantity strongly suggest its zero value, at least in lower mortality groups. Even at advanced age female survivability in 1999 Japan was [7] 97% till 51 year, 95% till 58 y, 85% till 73 y, 73% till 80 y (cf 6% in 1950). Yeast mortality was zero during half of its mean life span [3]. Similarly, only 2 (out of 7500) dietary restricted flies died at 8 days [2]. None of 1368 nematodes with changes in small number of their genes and tissues [5] died till 25 days (90 years in human terms), only 5% died till 40 days, and only 15% died in the first 3 months (over 200 "human years").

According to Eq. (1), at any age echelon mortality is as reversible as "infant" mortality, with the relaxation time ~1 year for humans, 1 day for flies, etc. So, it has short memory of the previous life history, and rapidly (within few percent of the life span) adjusts to current living conditions. Indeed, following unification of East and West Germany, within few years mortality in the East declined toward its levels in the West, especially among elderly, despite 45 years of their different life histories [14]. Dietary restriction resulted in essentially the same robust increase in longevity in rats [1] and decrease in mortality in Drosophilas [2], whenever restriction was switched on, i.e. independent of the previous "dietary" life history. However, when dietary restriction changes to full feeding, longevity remains higher than in the control group of animals fully fed throughout life. Also, when fly temperature was lowered from 27 C to 18 C or vice versa, the change in mortality, driven by life at previous temperature, persisted in the switched flies compared to the control ones. Such long memory of the life history may be related to insufficiently slow changes in temperature or feeding. It calls for comprehensive tests of short mortality memory in, and thus of rapid compared to life span mortality adaptation to, such conditions. Following the decrease in infant mortality with improving (medical and biological included) conditions, mortality of a homogeneous cohort may be reset to its value at a much younger age. Indeed, mortality of the female cohort, born in 1900 in neutral Norway, beyond 17 years of age monotonically decreased till 40 when it reversed to its value at 12 years. Then it little changed till 50 years. Only at 59 it restored its value at 17 years, i.e. 42 years younger. (The cohort probability $q(x)$ to die at any age x is calculated according to ref. 7 procedures and data). Dietary restriction, switched on day 14 [4], in 3 days restored Drosophila mortality at 7 days, i.e. 2.5 times younger. Thus, under certain conditions, predicted short memory and reversal of mortality to much younger age are observed in flies, rats, and humans; vanishing and very low mortality is seen in yeast, nematodes (including biologically amended ones), flies, and humans. This is inconsistent [9] with all evolutionary theories [10] of mortality.

Start quantitative verification of the universality of Eqs. (1) and (2) with two comments. Demo- and biodemographic data present age in years for humans, days for flies, etc. A non-stationary $q(0)$ is close to infant mortality if it changes relatively little within such time. This defines "regular" (in contrast to "irregular") conditions. They allow for the change in infant mortality which is very large (~50-fold [7]) and rapid on the life span scale. Generic inaccuracy of d(x) data is $\sim 1/D^{1/2}$ where $D$ is the number of deaths in the population of a given age in given conditions. When demographic fluctuations are consistent with this inaccuracy, denote the population as "well protected".

Mortality, especially in diverse conditions, is by far the best quantified for humans. This allows for comprehensive test of Eqs. (1,2). Equation (2) implies piecewise linear d(x) vs q(0). Populations in 18 developed countries over their entire history [7] (except for the years during, and immediately after, major wars, epidemics, food and water contamination, etc.), are well protected and regular. Over 3000 their male and female curves of d(x) vs the same calendar year q(0) may be approximated with several linear segments. (Further increase in the number of segments little changes the relative mean squared deviation from experimental curves). Figure 1, which is characteristic of all humans, demonstrates empirical d(60), d(80) and d(95) vs. q(0) and their piecewise linear approximations for Japanese females [15].

The number of parameters in Eq. (2) depends on the heterogeneity of the population. Consider the case when a heterogeneous population is distributed at two, e.g., the 1-st and 2-nd, intervals with the concentrations $c_1$ and $c_2 = 1 - c_1$ correspondingly. Then

$$q(0) = c_1 q_1(0) + (1-c_1) q_2(0); \quad d(x) = c_1 d_1(x) + (1-c_1) d_2(x) \tag{5}$$

By Eqs. (1, 2, 5), $q_1(0) = \alpha_1 q(0)$, $q_2(0) = \alpha_2 q(0)$, where

$$c_1 = (b_2-b)/(b_2-b_1); \quad \alpha_1 = (a-a_2)/[c_1(a_1-a_2)]; \quad \alpha_2 = (a_1-a)/[(1-c_1)(a_1-a_2)]. \tag{6}$$

The crossover to the next non-universal segment occurs when, e.g., $q_1(0)$ reaches the intersection $q_2 = (b_2 - b_1)/(a_1 - a_2)$ of the first and second universal segments in Eq. (1). Then $q_1(0) = q_2$ implies, by Eqs. (1) and (6), that $d^I(x) = a_1 q^I(0) + b_1$ [a subscript

*I* denotes an intersection in Eq. (2)]. So, by Eq. (1), this intersection falls on the first universal linear segments or its extension. Such universality is the criterion of the population distributed at two universal segments. Remarkably, demographic data [7] demonstrate that, except for few irregular years, this is the case in most developed countries (e. g., 1948–1999 Austria, 1921–1996 Canada, 1921–2000 Denmark, 1841–1898 England, 1941–2000 Finland, 1899–1897 France, 1956–1999 West Germany, 1906–1998 Italy, 1950–1999 Japan, 1950–1999 Netherlands, 1896–2000 Norway, 1861–2000 Sweden, 1876–2001 Switzerland)-see examples in Fig. 2. The intersections of their d(*x*) vs d(0) piecewise linear approximations determine 5 echelons of the universal (i.e. the same for all countries, thus for all humans) law, presented in Fig. 2. A general case (when the population is distributed at more than two universal segments) is more complicated, and may refine Fig. 2, but it also reduces to the universal law and the echelon fractions. The extrapolation of the universal d(60) and d(80) in Fig. 2 to $q(0) = 0$ yields d(60) = d(80) = 0, consistent with zero mortality till 80 years [9]. Similar study [9] demonstrates that species as remote from humans as protected populations of flies also yield the same (when properly scaled with the fly values of $q_j$) universal law, possibly with a different number of echelons. (See [17] for more details).

**4. Conjectures and challenges.** Human maximal lifespan, which remains ~120 years since ancient Rome (where birth and death data were mandatory on the tombstones) to present time, implies human maximal mean life span [16]. Meanwhile, a small number of perturbed genes and tissues increased mean life span in nematodes 6-fold (to 430 years in human terms), with no apparent loss in health and vitality. (One wonders how their cumulative damage, e.g., mutation accumulation, is eliminated). Universality implies this must be true in all animals, single cell yeast included, and suggests that universal mortality is in fact a disposable evolutionary byproduct. Equation (4) presents certain phase equilibrium as its possible mechanism, whose scaling parameters are related to biology and genetics of (possibly specific) cells. Clearly, physicists may be best qualified to test this mechanism in the case of, e.g., single cell yeast; verify the universality of its mortality law; if necessary, to refine the law with more additive parameters and estimate the contribution of non-universal mortality (for humans this may be done with existing life tables); to develop a microscopic model of universal mortality and transformation of a multitude of external and internal parameters into its scaling parameters; to establish the nature of these parameters. (For more details see [17]).

**5. Conclusions.** In evolutionary unprecedented protected populations mortality is dominated by the universal law, which implies accurate, reversible, rapid, stepwise mortality adjustment to current environment, and suggests that universal mortality is an evolutionary byproduct which may be eliminated. Indeed, Kenyon et al [5] increased mean life span of nematodes six-fold (to 430 years in human terms) with no apparent loss in health and vitality. The mechanism which is common to all animals, from humans to single cell yeast, reduces mortality to processes in a cell. Equations (3, 4) suggest adiabatically changing phase equilibrium. The transformation of multiple environmental factors into phase concentrations and universal equations of state remains a challenge and calls for extensive physical and biological study. Phase boundaries manifest and quantify the "rungs" in the universal "ladder" of mortality adaptation to extrinsic and intrinsic changes.

Aging may be addressed [5] by examining level of activity of surviving animals, quantifying their dynamics, and studying correlation with, and relation to, the universal mortality. Presented approach is applicable to other quantifiable biological phenomena.

**Acknowledgments.** I am very grateful to I. Kolodnaya for assistance. Financial support from A. von Humboldt award and R. & J. Meyerhoff chair is highly appreciated.

**Figure Captions**

Fig. 1. The period probabilities for live newborn 1950-1999 Japanese females to die between 60 and 61 (open squares), 80 and 81 (triangles), 95 and 96 (diamonds) years of age vs. infant mortality $q(0)$. Their relative mean squared deviations from piecewise linear approximations (straight lines) are 2.4%, 2.3% and 10%.

Fig. 2. Universal law for d(80) and d(60) (upper and lower curves, thick lines) vs. $q(0)$. Note that d(80) = 0 and d(60) =0 when $q(0)=0$. Diamonds and squares exemplify intersections of non-universal linear segments for (from left to right) England (two successive intersections), France, Italy and Japan, Finland, Netherlands, Norway, Denmark, France, England. Thin lines extend the universal linear segments.

Demography **34**, 1 (1997)), but with inadequate mathematical tools. Equation (1) demonstrates that universal dependence is explicit for d(x) vs q(0) in a single echelon.

14. Vaupel, J. W., Carey, J. R., Christiansen, K., Science **301**, 1679 (2003)

15. Until $\sim$ 65 years, d($x$) decreases when $q$(0) increases. Beyond $\sim$ 85 years, d($x$) increases together with $q$(0). In between, d($x$) exhibits a well pronounced maximum (smeared by generic fluctuations). Consider the origin of such dependence of d($x$)=$l$($x$)$q$($x$) on age $x$. When living conditions improve, the probability $l$($x$) for a newborn to survive to x increases, while the probability $q$($x$) to die at $x$ decreases. In young age the probability to survive to x is close to 1, d($x$) is dominated by $q$($x$), and thus monotonically decreases together with $q$(0). For sufficiently old age, low probability to reach $x$ dominates. It increases with improving living conditions, i.e. with decreasing $q$(0), thus d($x$) increases with decreasing $q$(0). At an intermediate age, when improving living conditions sufficiently increase survival probability, d($x$) increase is replaced with its decrease. Then d($x$) has a maximum at a certain value of $q$(0). Further study may yield the new lowest mortality echelon, which will dominate future mortality and its law, provide better statistics in old age, and possibly imply the d($x$) maximum at 95 and even more years of age.

16. It was first suggested by Strehler B.L.**.** Mildvan A.S., Science 132, 14 (1960). Its estimates see in Azbel' M.Ya., Proc. R. Soc. Lond. B **263**, 1449 (1996); Carnes, B. R., Olshansky S. J., Grahn, D. Biogerontology 4, 31 (2003), and [14].

17. Azbel', M. Ya. http://arxiv.org/abs/q-bio.QM/0403007

FIG. 1

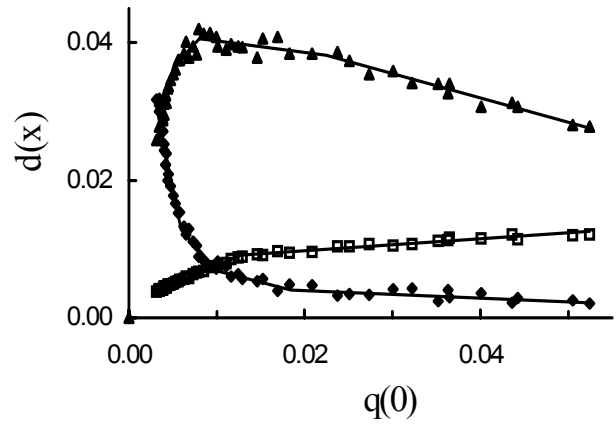

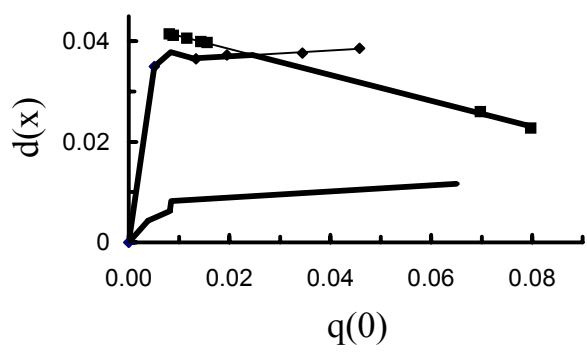